\documentclass[a4paper]{jpconf}
\usepackage{graphicx}
\newcommand{\cor}[1]{\left\langle{#1}\right\rangle}
\newcommand{\zt}{\tilde{z}}
\newcommand{\f}{\frac}
\newcommand{\om}{\omega}
\newcommand{\rsq}{{{\cal R}^2}} 
\begin{document}
\title{Quark-Gluon Plasma/Black Hole duality from
Gauge/Gravity Correspondence}

\author{Robi Peschanski}

\address{Service de Physique Th\'eorique, CEA Saclay, France}

\ead{robi.peschanski@cea.fr}

\begin{abstract}
The Quark-Gluon Plasma (QGP) is the QCD phase of matter  expected 
to be formed at small proper-times in the collision of heavy-ions at high energy. 
Experimental observations seem to favor a strongly coupled QCD plasma with the 
hydrodynamic properties of a quasi-perfect fluid, \emph{i.e.} $\!$ rapid 
thermalization 
(or isotropization) and small viscosity. The theoretical investigation of 
such properties is not obvious, due to the the strong coupling. The Gauge/Gravity 
correspondence provides a  stimulating framework to explore the strong coupling 
regime of gauge theories using the dual string description. After a brief 
introduction to Gauge/Gravity duality, and among various existing studies, we 
focus 
on challenging problems of  QGP hydrodynamics, such as viscosity and 
thermalization,  in terms of gravitational duals of both the static and  
relativistically evolving plasma. We show how a Black Hole geometry arises 
naturally from the dual properties of a nearly perfect fluid and  explore the 
lessons and prospects one may draw for actual heavy ion collisions from the  
Gauge/Gravity duality approach. 
\end{abstract}

\section{Introduction}
\begin{figure}
[hb]
\begin{center}
\includegraphics[width=22
pc]{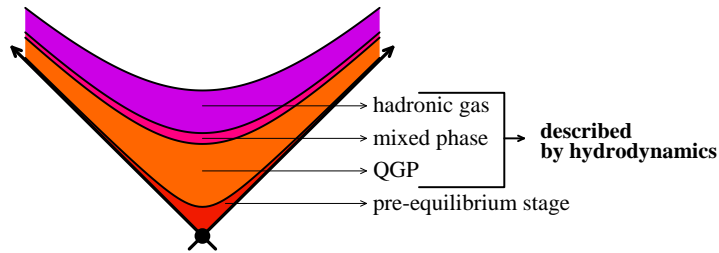}
\end{center}
\caption{\label{QGP}Description of QGP formation in heavy ion collisions.}
\end{figure}

The formation of a QGP (Quark Gluon Plasma) is expected to be realized in 
high-energy 
heavy-ion collisons, {\it e.g.} at RHIC and soon at the LHC. One of the main tools
for 
the description of such a formation is the relevance of relativistic hydrodynamic 
equations in some intermediate stage of the collisons, see Fig.\ref{QGP}.
The problem of the hydrodynamic description is the uneasiness of the   relation 
with the 
underlying 
fundamental theory. Indeed, the experimental observations seem to indicate an 
almost 
perfect-fluid behaviour with small shear viscosity, which naturally leads to 
consider a 
theory at strong coupling and thus within the yet unknown nonperturbative regime 
of 
QCD. Moreover the QGP formation appears to be fast, which may
also 
point towards strong coupling properties. Another key point of the standard 
description is the approximate boost-invariance of the process as predicted in the 
seminal paper of Ref.\cite{Bjorken:1982qr}. The goal of the string theoretic 
approach is 
to make use 
of the gauge/gravity correspondence as 
a 
way to tackle the problem of the hydrodynamic behaviour from the fundamental 
theory 
point of view. It allows to draw quantitative  relations
between a strongly coupled gauge field theory and a weakly coupled string theory

More specifically, the AdS/CFT correspondence between the ${\cal N}=4$ 
supersymmetric 
$SU(N)$ gauge theory and  superstrings in 10 dimensions can be used as a 
calculational 
laboratory for this kind of approach, at least as a first stage before a more 
realistic 
application to QCD. The unconfined character of the QGP gives some hope that the 
explicit AdS/CFT  example could be useful despite
the lack of asymptotic freedom and other aspects specific of QCD. 
\section{Gauge/Gravity correspondence}
As well-known, superstring theory is defined as the quantum embedding of a 
2-dimensional 
world sheet describing the moving string into a $10$-dimensional space. In a  
qualitative way, the gauge-gravity correspondence can be given a 
geometric 
interpretation \cite{Schomerus:2007ff}. 
\begin{figure}
[hb]
\begin{center}
\includegraphics[width=18
pc]{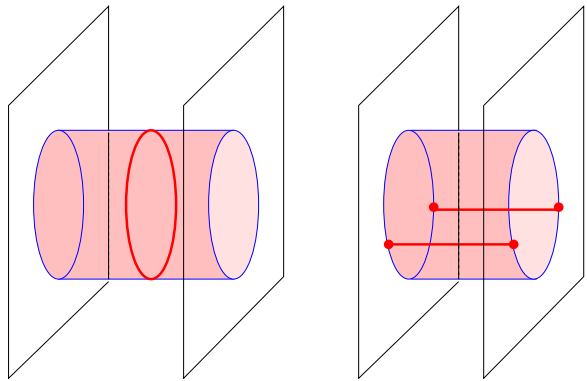}
\end{center}
\caption{\label{open}Open $\Leftrightarrow$ closed 
string duality (from Ref.\cite{Schomerus:2007ff}).}
\end{figure}
Considering  the cylinder stretched between the two $branes$ in Fig.\ref{open}, it 
may 
be described as a propagating closed string (Fig.\ref{open} left) or an open 
string 
with ends on 
the 
branes performing a loop (Fig.\ref{open} right). In terms of  fields, a  gravity 
interaction 
between 
branes (left) becomes equivalent to  a 1-loop gauge interaction (right). This 
gauge/gravity 
$duality$ is also a weak/strong coupling one. Indeed, considering a long distance 
between branes, it corresponds to a weak  and thus $classical$ gravitational 
interaction 
between branes (left), while in gauge theorical terms, it is expected to imply a 
long-distance gauge field interaction including many excitation modes of the 
1-loop 
open 
string, and thus a $quantum$ field theory at strong-coupling (right). This duality 
property can be given a quantitative framework, in particular in the case of 
AdS/CFT 
correspondence. 
\section{Holography and Hydrodynamics}
One typical and fascinating aspect of the gauge/gravity duality is the property of 
$holography.$ It states that the amount of information contained in the boundary 
gauge 
theory (on the brane) is the same as the one contained in the bulk string theory. 
In our 
problem, we shall make use in a quantitative way of this property by taking 
advantage of 
one of the remarkable relations due to the holographic renormalization
\cite{Skenderis:2002wp}. One can write
$$
g_{\mu\nu}=g^{(0)}_{\mu\nu} 
{(=\eta_{\mu\nu})}+z^2 
g^{(2)}_{\mu\nu} 
{(=0)}+z^4 
\cor{T_{\mu\nu}} + 
z^6 g^{(3)}_{\mu\nu} {\ldots} + \ ,
$$
where $g_{\mu\nu}$ is the bulk metric in 5 dimensions, $\eta_{\mu\nu},$ the 
boundary 
metric in physical (3+1)  Minkowski space and $\cor{T_{\mu\nu}},$ the v.e.v. of 
the 
physical energy-momentum tensor. The higher coefficients of the expansion over 
the 
fifth dimension $z$ can be obtained by the Einstein equations in the bulk provided 
the 
boundary  energy-momentum tensor fulfils the  zero-trace and continuity equations.

The interesting observation  on which we shall elaborate, namely that there is a 
nontrivial dual relation between a perfect fluid at rest in (3+1) dimensions and  
a 
static 5d black hole in the bulk \cite{Balasubramanian:2001nb} can be proven using  
holographic renormalization. Indeed, implementing the perfect 
fluid 
behaviour with diagonal elements $\cor{T_{\mu\nu}}\equiv  {\rm diag} 
\{\epsilon,p_1,p_2,p_3\} \propto {\rm diag} \{3,1,1,1\},$ one can resum 
\cite{Janik:2005zt} the 
holographic 
expansion and get ($via$  a change of variables $z\!\to\!\zt$) the following bulk 
metric
$$ 
ds^2=-\f{1-\zt^4/\zt_0^4}{
\zt^2} dt^2
+\f{dx^2}{\zt^2}+\f{1}{1-\zt^4/\zt_0^4
} \f{d\zt^2}{\zt^2}\ ,
$$ 
where one recognizes the Black Hole (in fact a black brane) with a static  horizon 
at 
$\zt_0$ in 
the 5th dimension. In fact there exists a one-to-one correspondence between the 
thermodynamic properties of the  Black Hole (BH) and those of the perfect fluid 
(PF), 
namely its {temperature} ($ T_{BH} =  
 \epsilon^{\f{1}{4}} = T_{PF}$) and {entropy} ($  S_{BH} \sim Area 
= \epsilon^{\f{3}{4}}=S_{PF}$).

It is in this context of a static Black hole configuration that one can go further 
than 
the perfect fluid approximation and derive the viscosity \cite{Policastro:2001yc}, 
using 
the Kubo formula. Indeed, the duality properties extend to a relation between the 
correlators of the energy-momentum tensor in two  space-time points  at zero 
frequency 
$\om=0$ and the absorption cross section of a graviton by the static Black Hole in 
the 
bulk. One writes
$$  \sigma_{abs}(\om)\propto \int d^4x\ \f 
{e^{i\om t}}{\om} \ 
{\cor{\left[T_{x_2x_3}(x),T_{x_2
x_3}(0)\right]}}
\Rightarrow 
\f{\eta}{S}\equiv \f 
{\sigma_{abs}(0)/16 \pi G}
{A/4 G}=\f 1{4\pi}\ ,
$$ 
where $S=S_{BH}\equiv A/4 G$ is the famous entropy-area relation of a Black Hole. 
From this 
relation, and putting numbers, it appears that the viscosity is weak, much weaker 
than 
the one computed in the weak coupling regime and eventually realizing an absolute 
viscosity 
lower 
bound.

\section{QGP and Black Holes: From Statics to Dynamics}

\begin{figure}
[ht]
\begin{center}
\includegraphics[width=18
pc]{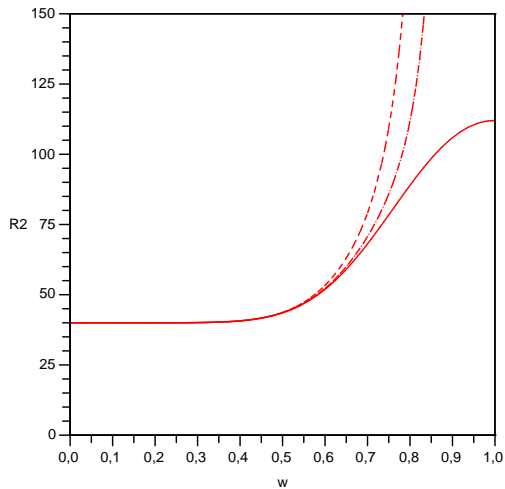}
\end{center}
\caption{\label{ricci} The curvature scalar  $\rsq$ for $s = \f 43 {\pm .1}\ :$ 
Nonsingular 
Geometry $\Leftrightarrow$ Perfect Fluid.}
\end{figure}
The previous results were obtained for static configurations, $i.e.$ for a 
thermalized QGP at rest. In order to take into account, as much as possible, the 
actual 
kinematics 
of a heavy-ion collision, it is required to introduce the proper-time expansion of 
the 
plasma. On the gravity side, it calls for studying non-equilibrium geometries, 
eventually 
of 5d Black Hole configurations, which represent in itself a nontrivial and 
interesting 
issue.

In Ref.\cite{Janik:2005zt}, it was proposed to build the dual geometries of the 
standard 
Bjorken flow \cite{Bjorken:1982qr}, that is the description of a boost-invariant 
expansion 
of the QGP, which is expected to correspond to the physical situation in the 
central 
rapidity region of the collision. In this context the questions why the QGP fluid 
appears 
to be nearly perfect (small viscosity) and why its thermalization time can be 
short  have been discussed
among 
other problems. 

One starts from a family of proper-time dependent, boundary energy-momentum 
tensors
$$\cor{T_{\mu\nu}}\equiv  {\rm diag} \left\{\scriptstyle{f(\tau)\ ,\ -\tau^3 
\f{d}{d\tau} 
f(\tau)\!-\!\tau^2 f(\tau)\ ,\ f(\tau)\!+\! \f{1}{2}\tau 
\f{d}{d\tau} f(\tau)\ ,\ f(\tau)\!+\! \f{1}{2}\tau 
\f{d}{d\tau} f(\tau)}\right\},$$
where the function $f(\tau)\propto \tau^{-s}$ corresponds to an interpolation 
between 
different relevant regimes, namely perfect fluid $(p_1\!=\!p_{2,3})$ $s\!=\!-\!\f 
43,$  
free streaming $(p_1\!=\!0)$ $s\!=\!-\!1,$ fully anisotropic  $(p_1\!=\!-p_{2,3})$ 
$s=0.$
Using the holographic renormalization to compute the coefficients of  the 
corresponding 
metrics in the expansion on the fifth dimension and after resummation, it was 
possible to 
solve the dual geometry for given  $s$ at asymptotic  proper-time $\tau.$ It 
reveals 
the 
existence of a scaling property of the solutions in terms of  the proper-time 
dependent variable  
$v=\f{z}{\tau^{1/3}}\ .$

Analyzing the family of solutions as a function of $s,$ it appears that the only 
nonsingular solution for invariant scalar quantities (here the square of the Ricci 
tensor
$\rsq=R^{\mu\nu\alpha\beta}R_{\mu\nu\alpha\beta},$ see Fig.\ref{ricci}), is 
obtained for $s=4/3.$ This solution is the only one of the family corresponding  to 
a Black Hole 
moving 
away in 
the fifth dimension. Hence 
the perfect-fluid case is 
singled out  and the  moving Black Hole in the bulk corresponds through duality  
to the expansion of the QGP taking place in the 
boundary. 
Consequently, the BH horizon moves as $z_h(\tau) \propto
 \tau^{\f{1}{3}},$
 the temperature as $T(\tau) \sim  
{1}/{z_h} \sim \tau^{-\f{1}{3}},$ and the entropy stays constant since $ S(\tau) 
\sim 
Area 
\sim \tau \cdot {1}/{z_h^3} \sim 
const.$ Note that again the physical thermodynamical variables of the QGP are the 
same as 
those one may attribute to the BH in the bulk (with the reservation that 
thermodynamics of a moving BH may rise   nontrivial interpretation problems). 
Hence 
one finds a 
concrete 
realization of the idea \cite{Nastase:2005rp} of a duality between the QGP 
formation and a 
moving  a moving Black Hole.

\section{Thermalization}
There has been a 
lot of activity along the lines of the AdS/CFT correspondence 
and 
its 
extensions to various geometric configurations. Dual studies of quark energy loss 
\cite{Herzog:2006gh},
 jet quenching \cite{Liu:2006ug}, quark
dragging \cite{Gubser:2006bz}, etc... have been and are being performed. Sticking 
to the configurations 
corresponding to  an expanding plasma and going beyond the first order terms in 
proper-time, one has obtained results \cite{Janik:2006ft} on the viscosity, 
confirming the 
universal bound of Ref.\cite{Policastro:2001yc}, on the relaxation time of the 
plasma and 
very recently on the inclusion of flavor degrees of freedom \cite{Gro[]e:2007ty}.

Let us finally focus on the thermalization problem, which can be usefully taken up 
using 
the gauge/gravity duality in the strong coupling hypothesis. the problem is to 
give 
an 
explanation to the strikingly small thermalization time required for the formation 
of a 
QGP as can be abstracted from the experimental observations. Already in 
ref.\cite{Janik:2006gp}, it has been found that if one performs a small deviation 
from the 
Black hole metric by coupling with a scalar field and analyze the so-called 
quasi-normal 
modes defining the way how the system relaxes to its initial state, one finds a 
numerically small value of the relaxation time in units of the local (and 
evolving) 
temperature. Even if a definite value of this  relaxation time cannot be inferred 
at this stage due to  scale-invariance, this result was suggestive of a stability  
of the QGP  in the strong coupling regime
with 
respect to perturbations out of equilibrium.

In order to go further, one has to deal with the problem of the QGP evolution at 
small 
proper-times. In Ref\cite{Kovchegov:2007pq}, the holographic renormalization 
program has 
been pursued for the small proper-time expansion. Relaxing the selection of the 
appropriate 
metric by requiring  only the metric tensor to be a real and single-valued 
function 
of 
the 
coordinates everywhere in the bulk, one finds an unique solution corresponding to 
the 
``fully anisotropic case'', $i.e.\ s=0.$ 
\begin{figure}
[h]
\begin{center}
\includegraphics[width=18
pc]{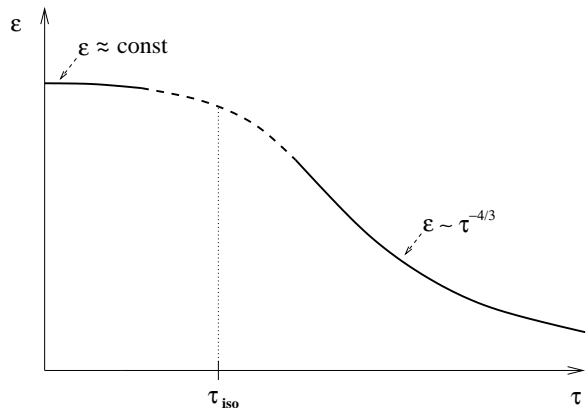}
\end{center}
\caption{\label{therma} Evaluation of  the isotropization/thermalization time 
(from 
Ref.\cite{Kovchegov:2007pq}).}
\end{figure}
In the same paper, an evaluation of the range of the isotropization time has been 
proposed, by extrapolation of realistic estimates abstracted from experiments to 
the supersymmetric case. The 
idea is to match the large and small proper-time regimes at some value of the 
proper-time  $\tau_{iso}.$ This proper-time is mathematically defined as the 
crossing value for the branch-point singularities of both regimes. Physically, it 
is expected to give an estimate of the proper-time range during which which the
medium evolves from the full anisotropic regime (small $\tau$) to the perfect 
fluid 
one (large $\tau$).
cd 
In order to give an idea of the possible physical implications of this strong 
coupling scheme, let us shortly reproduce the estimate made in 
Ref.\cite{Kovchegov:2007pq}.  Implementing the estimated physical value of the 
energy density at some proper-time
($e.g.\  \epsilon(\tau)\  = \ e_0\ 
\tau^{4/3}\vert_{\tau=.6} 
\sim 15\ {\scriptstyle GeV fermi^{-3}}$) one finds 
$$\tau_{iso}=\left(\f {3 
N_c^2}{2\pi^2e_0}\right)^{3/8}\sim \ .3\  
{\scriptstyle fermi}\ .$$ This short isotropization time thus seems a 
characteristic feature of the strong coupling scenarios. It is clear that more 
realistic estimates should take into account less idealized dual models, 
corresponding  to  QCD, such as the lack of supersymetry and 
the finite numbers of colors. However, the non confined character of the QGP and 
the robustness of some predictions (such as the $\eta/S$ ratio) may give some 
confidence that this short isotropization time could be a reasonable estimate at 
strong coupling.  
\section{Conclusions and outlook}
From the present rapid (and partial) survey of some of the results obtained in the 
AdS/CFT approach to the formation and expansion of the Quark-Gluon plasma in 
heavy-ion collisions, it appears that the gauge/gravity correspondence is a 
promising way to explore some features of QCD at strong coupling. Indeed some 
general features of  this correspondence, relating at long distances the closed 
and 
open string geometries (see Fig.\ref{open}) are expected to 
be valid in principle 
for various dual schemes  and thus, hopefully, QCD.

In practice, the quantitative dual schemes have been more precisely elaborated for 
the specific AdS/CFT case,  $i.e.$ the gauge theory with ${\cal N}=4$ 
supersymmetries. Among the results, it gives a 
calculable link between the hydrodynamic  quasi-perfect fluid behaviour on the 
gauge theory side with a BH geometry in the higher dimensional gravity side 
in and AdS background. This relation can be extended from the static case to a 
dynamical regime reflecting (within the AdS/CFT framework) the relativistic 
expansion of the corresponding quark-gluon plasma. This, and many other 
applications, some of them using more complex geometries, less supersymmetric 
backgrounds and examining other observables, gives hope on the fruitful  
possibilities of the gauge/gravity approach to the QGP formation.

As an outlook, it is worth mentionning some of the possible new directions of 
study
one is led to consider. Starting with the more technical ones,  it is known that 
the Bjorken flow is not exactly verified in heavy-ion collisions, since the 
observed 
distribution of particles is nearly gaussian  in rapidity and thus not reflecting 
exactly the  boost-invariance of the Bjorken flow. It would be interesting to 
investigate dual properties for non-boost invariant flows, such as the Landau 
solution \cite{Landau:1953gs}. On a more general  ground, the whole approach still 
concerns only the hydrodynamical stage of the QGP expansion. It would be important 
to attack both the initial (partonic) and final (hadronic) stages of the reaction 
in the same framework and thus the problem of {\it phase transitions} during the 
collision. Finally, one would like to have more realistic dual frameworks 
including 
a finite number of colors, flavor degrees of freedom and no (or broken) 
supersymmetry. 
 
\ack 
I would like to address special thanks to Romuald Janik for our stimulating and  
fruiful collaboration.

\end{document}